\documentclass[a4paper]{jpconf}
\usepackage{graphicx}
\usepackage{subfigure}
\begin{document}
\title{Excitation of baryonic resonances in stable medium-mass nuclei of Sn}
\author{
J L Rodr\'{i}guez-S\'{a}nchez$^{1,2}$,
J Benlliure$^1$,
E Haettner$^2$,
C~Scheidenberger$^{2,3}$,
J Vargas$^1$,
Y Ayyad$^1$,
H Alvarez-Pol$^1$,
J~Atkinson$^2$,
T~Aumann$^2$,
S~Beceiro-Novo$^1$,
K~Boretzky$^2$,
M~Caama\~{n}o$^1$,
E~Casarejos$^4$,
D~Cortina-Gil$^1$,
P~D\'{i}az Fern\'{a}ndez$^1$,
A~Estrade$^2$,
H~Geissel$^{2,3}$,
K~Itahashi$^5$,
A~Keli\'{c}-Heil$^2$,
H~Lenske$^6$,
Yu~A~Litvinov$^2$,
C~Paradela$^1$,
D~P\'{e}rez-Loureiro$^1$,
S~Pietri$^2$,
A~Prochazka$^2$,
T~R~Saito$^{2,5}$,
M~Takechi$^2$,
Y~K~Tanaka$^{2,3,5}$,
I~Vida\~{n}a$^7$,
H~Weick$^2$,
J~S~Winfield$^2$ for the Super-FRS experiment collaboration
}
\ead{joseluis.rodriguez.sanchez@usc.es}
\address{$^1$ University of Santiago de Compostela, E-15782 Santiago de Compostela, Spain}
\address{$^2$ GSI Helmholtzzentrum f\"{u}r Schwerionenforschung GmbH, D-64291 Darmstadt, Germany}
\address{$^3$ Justus-Liebig Universit\"{a}t Giessen, D-35392 Giessen, Germany}
\address{$^4$ Universidad de Vigo, E-36200 Vigo, Spain}
\address{$^5$ RIKEN Nishina Center, Saitama 351-0198, Japan}
\address{$^6$ Institut f\"{u}r Theoretische Physik der Universit\"{a}t, D-35392 Giessen, Germany}
\address{$^7$ INFN, University of Catania, I-95123 Catania, Italy}
\begin{abstract}
Isobaric charge-exchange reactions induced by beams of $^{112}$Sn have been investigated at the GSI facilities using the fragment separator FRS.
The high-resolving power of this spectrometer makes it possible to obtain the isobaric charge-exchange cross sections with an accuracy 
of $3\%$ and to separate quasi-elastic and inelastic
contributions in the missing-energy spectra, in which the inelastic component is associated
to the in-medium excitation of baryonic resonances such as the $\Delta$ resonance. 
We report on the results obtained for the $(p,n)$ and $(n,p)$ channels excited by using different targets that cover a large range in neutron excess.
\end{abstract}

\section{Introduction}
The study of baryonic resonances in nuclear matter is considered as a natural extension of nuclear physics. 
It is well known that the structure of some baryons, such as the $\Delta$ resonance, and their excitation 
spectrum is one of the unsolved issues of strong interaction physics, but this also becomes an unsolved problem in nuclear physics 
when the resonances are excited in nuclear matter. 
The investigation of the excitation in the nuclear medium of these resonances is very important to understand 
the pion production in heavy-ion collisions \cite{Aoust2006,JLRS2018}, 
because their main decay channel is going into one nucleon and one or more pions, 
as well as for the understanding of three body forces \cite{krebs} or the not fully
solved problem of the missing Gamow-Teller strength~\cite{liang}. 
Moreover, the early appearance of $\Delta$-isobars in dense nuclear matter has also inspired many studies relevant to 
neutron stars \cite{feng2016,sen19}, in particular, 
very compact stellar configurations are reached due to the introduction of $\Delta$-isobars \cite{drago14,drago16}. 
Recently though, different works have shown that the properties (mass and radius) 
of these stellar configurations depend much more on how the mass of the $\Delta$-isobars changes with the nuclear density \cite{cai15}.

Unfortunately, the in-medium properties of baryonic resonances are not well understood. Up to now, the in-medium effects of $\Delta$-isobars 
have been studied in heavy-ion collisions at kinetic energies above the production threshold using pion-nucleus and 
nucleon-nucleus reactions in direct kinematics \cite{Kaufman,chiba91}, 
where there is not a good control over the production of residual fragments and, thus, over the number of collisions. 

In this work, for the study of baryonic resonances in nuclear matter, we use isobaric charge-exchange reactions induced by 
relativistic beams of stable medium-mass projectile nuclei. These reaction channels guarantee that with a large probability 
the charge-exchange process is mediated by a quasi-free nucleon-nucleon collision \cite{bachelier}. In the case of inelastic collisions 
the excited baryon resonances, produced in the overlapping region between projectile and target nuclei, decay by pion 
emission. To preserve the initial number of nucleons in the projectile remnant, pion decay should not induce any sizable 
excitation energy in the projectile.

In this proceeding we will describe the experiment that we performed at GSI using beams of $^{112}$Sn to induce isobaric charge-exchange reactions. 
The atomic and mass numbers of the projectile residues as well as their momentum distributions were obtained by using 
the zero-degree magnetic spectrometer fragment separator (FRS) \cite{Geissel1992}. 
The momentum resolution of this spectrometer was already proved some time ago 
to be sufficient to clearly identify the quasi-elastic (Gamow-Teller, spin-isospin dipole and quadrupole resonances, etc) and inelastic (baryon excitations) 
components in the missing-energy spectra of nuclear residues produced in isobaric charge-exchange reactions, 
which were induced by projectiles of $^{208}$Pb \cite{kelic2004}. Though the momentum spectra were obtained with 
a poor resolution due to the large angular and energy stragglings suffered by the projectiles and nuclear residues 
in the layers of matter placed along the beam line. To go further, we decreased these stragglings by reducing 
the layers of matter in beam and also improved the position resolution of the detector used to measure the momentum of the ejectiles. 
In the following, we will detail the experimental technique and setup and then we will present some results.

\section{Experiment}

The experiment was performed at the GSI accelerator facilities in Darmstadt (Germany) using the SIS18 synchrotron combined with 
the fragment separator FRS. Beams of $^{112}$Sn at 1$A$ GeV were delivered by the SIS18 synchrotron with an intensity around 
10$^{8}$ ions per spill and were then guided up to the FRS entrance to produce the isobaric charge-exchange reactions 
in different thin targets of around 100 $mg/cm^{2}$.

The FRS is a zero-degree high-resolving power spectrometer with a typical resolving power 
of $B\rho/\Delta B\rho =1500$ and with an angular acceptance 
of $\pm$15 mrad around the central trajectory. In the present work, the FRS spectrometer is used in the achromatic mode, where 
the reaction target is placed at the FRS entrance and the full spectrometer 
is then utilized to separate and identify the nuclear residues, as shown in Fig. 1(a). The nuclear residues 
are identified by measuring their B$\rho$, velocity and energy loss by using high resolution time-projection chambers (TPC) for tracking (300 $\mu$m FWHM), 
scintillator detectors for time-of-flight and
multi-sampling ionization chambers (MUSIC) for energy loss. These measurements also allow us for the accurate
determination of the longitudinal momenta of the reaction products. 
With these measurements, we can define two important observables 
for the characterization of the isobaric charge-exchange reaction, the cross section of the process and the
ejectile missing-energy spectra. 

Isobaric charge-exchange cross sections were accurately determined 
by normalizing the production yield of the charge exchange residual nuclei to 
the number of projectiles and target nuclei. The number of incoming projectiles was obtained by using 
a secondary electron monitor (SEETRAM) placed at the entrance of the FRS. 
The uncertainty in the determination of the incoming projectiles was around 3$\%$. 
The number of charge exchange residues is obtained from the corresponding identification matrix at the 
final focal plane by gating on the fragment of interest, as shown in Fig. 1(b). Details about detector calibrations 
and data corrections can be found in Ref.~\cite{JL2017v0}.

\begin{center}
\begin{figure}[h]
\begin{center}
\includegraphics[width=0.58\textwidth,keepaspectratio]{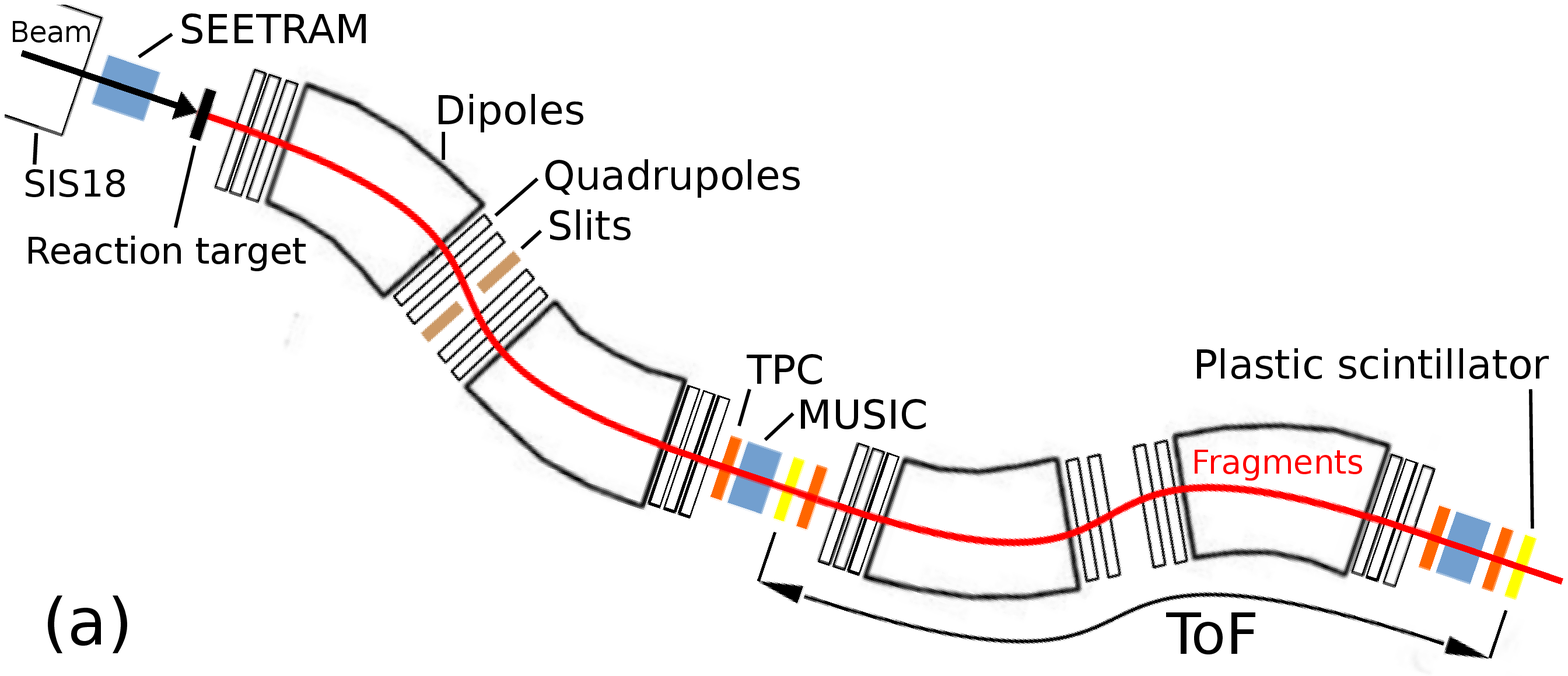}
\includegraphics[width=0.41\textwidth,keepaspectratio]{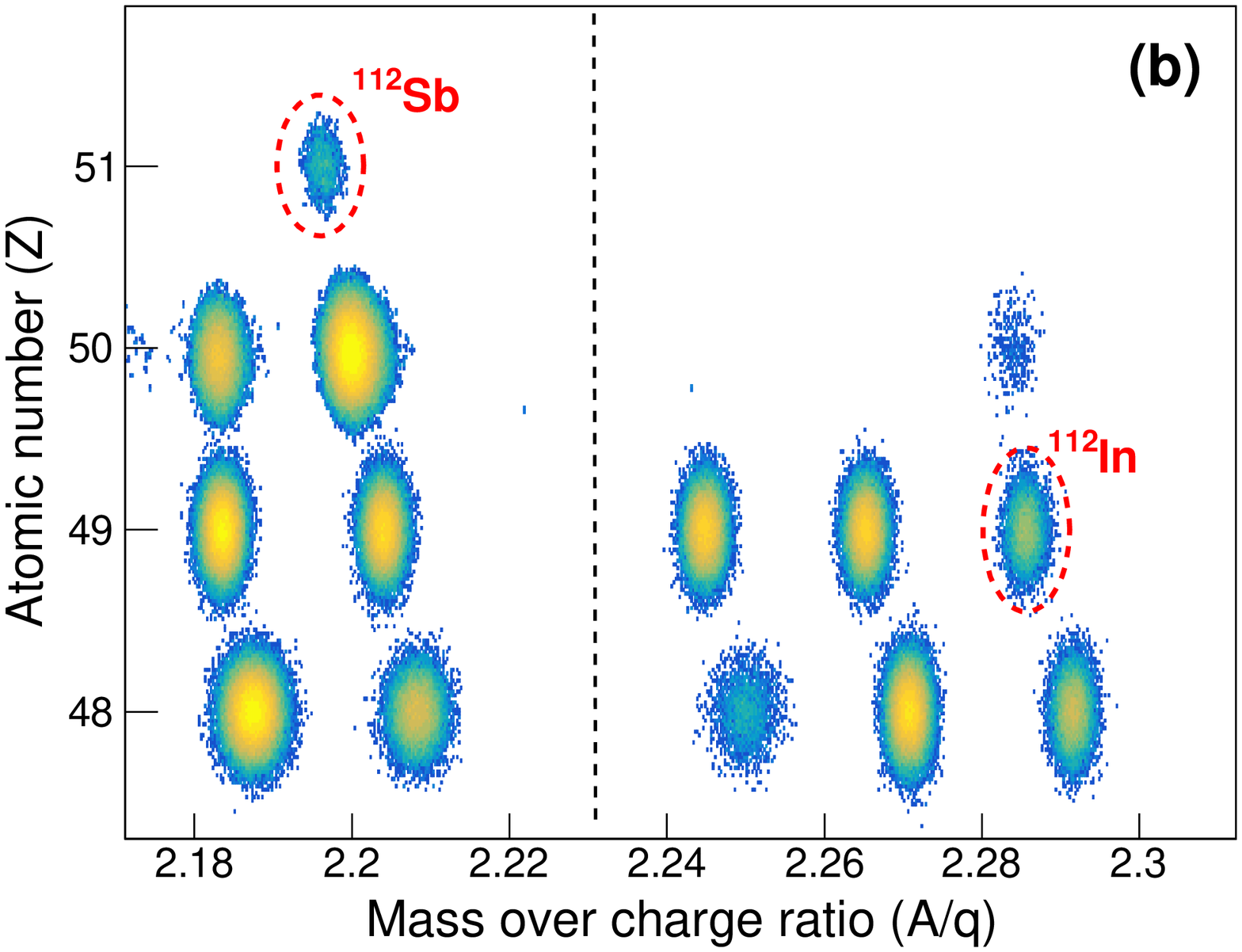}
\caption{(Color online) (a) Schematic view of the FRS experimental setup used in the present work with the reaction target at the entrance of the FRS spectrometer 
to induce isobaric charge-exchange reactions of stable tin isotopes. (b) Identification matrix at the final focal plane of the FRS obtained from reactions induced 
by $^{112}$Sn impinging on a carbon target. The figure was obtained by overlapping two magnetic settings of the FRS separated by a dashed line.
}
\end{center}
\end{figure}
\end{center}

The second observable that we use in this work is the missing-energy spectra of the recoiling 
charge-exchange residues. This observable was obtained from the measured longitudinal momenta of the residual
nuclei, which was also corrected by the dependence on the beam-extraction time given by the synchrotron SIS18~\cite{Tanaka18}.  
Thanks to the high-resolving power of the FRS together with a sizeable reduction of matter along the beam line we allowed to obtain 
this spectrum with a resolution of around 10 MeV (FWHM), which is a factor of 2 better than that was obtained at the Laboratoire National SATURNE in Saclay (France)~\cite{roy1988}. 
In addition, an unfolding procedure, based on the Richardson-Lucy method with a regularization 
technique to optimize the stability of the solution against statistical fluctuations~\cite{vargas}, was used to improve the sharpness of the spectrum.

\section{Results}

\begin{center}
\begin{figure}[h]
\begin{center}
\includegraphics[width=0.48\textwidth,keepaspectratio]{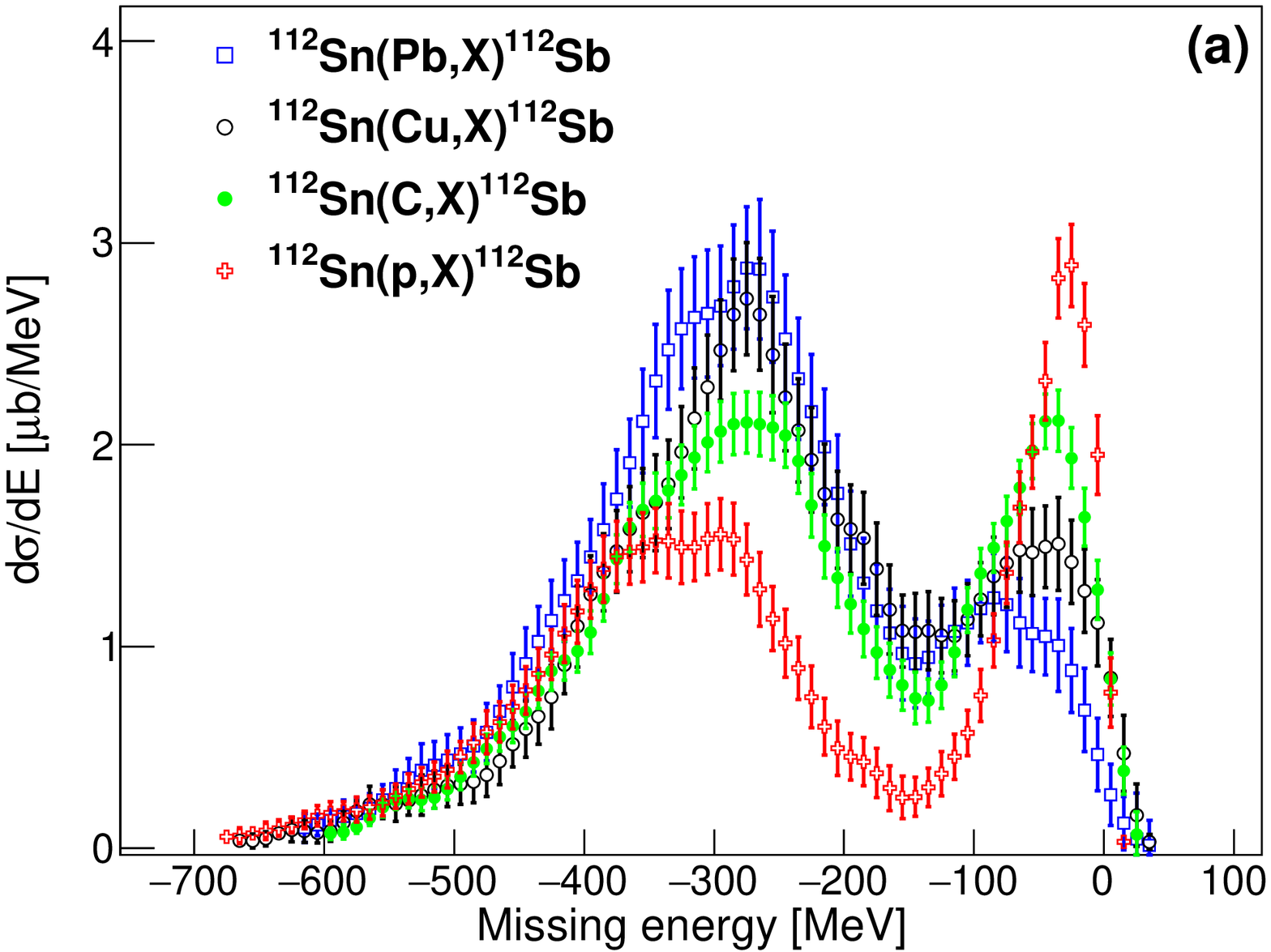}
\includegraphics[width=0.48\textwidth,keepaspectratio]{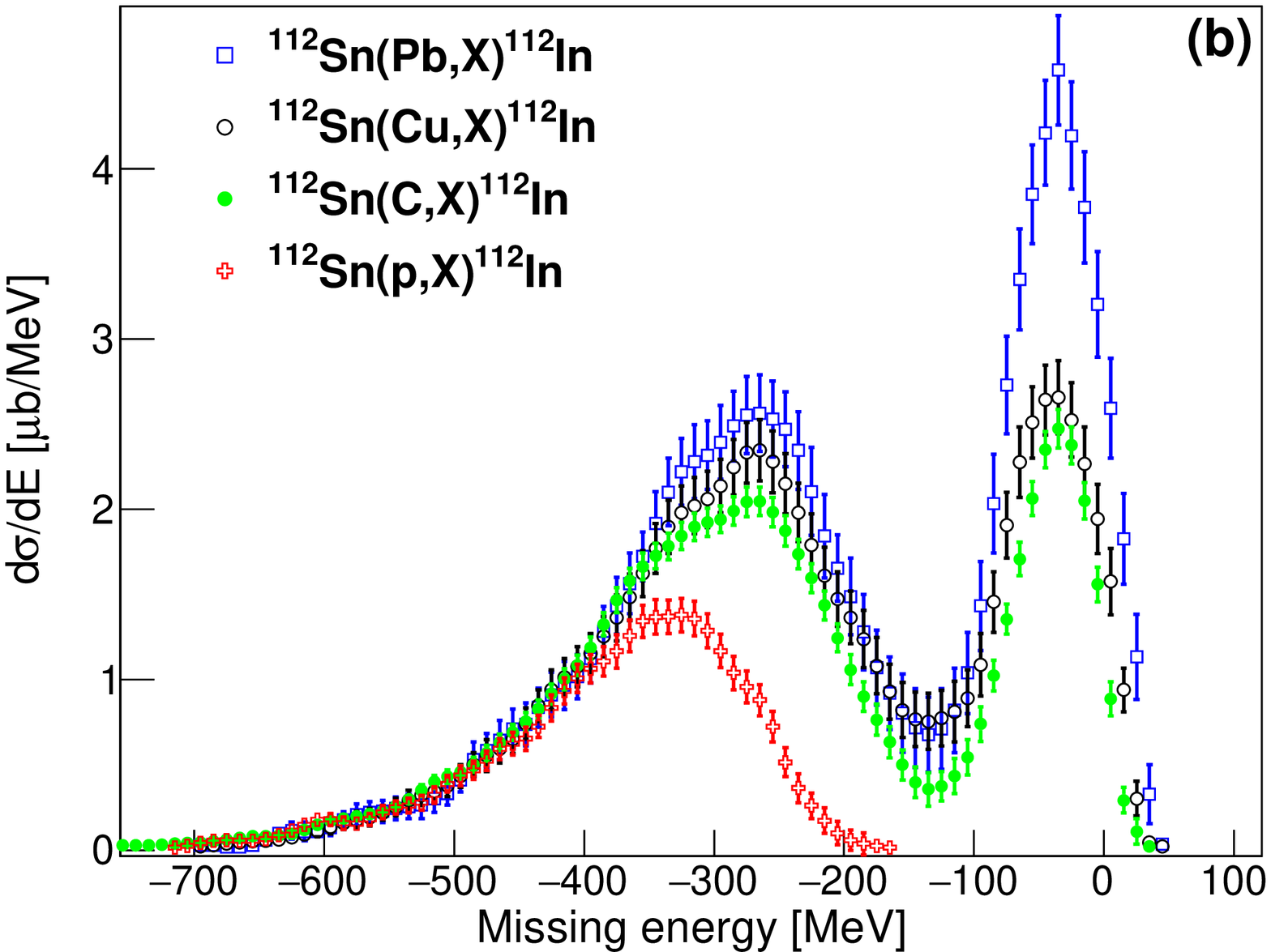}
\caption{(Color online) Missing-energy spectra for the $(n,p)$ (a) and $(p,n)$ (b) channels obtained from isobaric charge-exchange reactions 
induced by ions of $^{112}$Sn in different targets.
}
\label{fig:2}
\end{center}
\end{figure}
\end{center}

In Fig. \ref{fig:2} we show the missing-energy spectra obtained from the isobaric charge-exchange reactions 
induced by ions of $^{112}$Sn in different targets at projectile kinetic energies of 1$A$ GeV. 
These spectra show that the spin-isospin response of nuclei is concentrated in two energy domains.
At missing energies close to zero it corresponds to the excitation of particle-hole states: the $L = 0$ Gamow-Teller giant resonance, 
the $L = 1$ spin-isospin dipole and $L = 2$ spin-isospin quadrupole resonances, and also to a quasi-elastic charge exchange mechanism 
on a nucleon of the target and projectile. At lower missing energies, the observed peak in the spectrum around -300 MeV, which is also the difference 
between the nucleon mass and the $\Delta$ mass, corresponds to a nucleon being excited into a $\Delta$ resonance in the target or projectile. 
This twofold spectrum is a common feature of all isobaric charge-exchange reactions induced at high kinetic energies (more than 0.6$A$ GeV)
and illustrates clearly that the nucleon and the $\Delta$ resonance are two states of the same particle.

For the $(n,p)$ channel one can see that the cross section of the quasi-elastic peak decreases with 
the target size while the inelastic peak increases. In the case of the $(p,n)$ channel the cross section of both peaks increases 
with the target size, although this increase is faster for the quasi-elastic peak. This is also a clear proof of the competition 
between the formation of baryonic resonances and particle-hole states that could be used to extract information about the quenching factors 
of the Gamow-Teller transitions \cite{Horst18}.

Another interesting result is the effect due to the excitation of baryonic resonances in target and projectile. As demonstrated from the study of 
($^{3}$He,t) reactions at 2 GeV \cite{PFC1992}, there is a shift of 50-70 MeV between the $\Delta$ excitation in target and projectile 
due to the kinematics of these processes is different. 
For instance, in the $(p,n)$ channel one can see that for the proton target the maximum is around -330 MeV 
due to the excitation of a $\Delta^{++}$ in the target while in the case of C, Cu and Pb targets this maximum moves towards higher energies, 
observing an energy shift of around 60 MeV due to the excitation of $\Delta$ resonances in the projectile~\cite{PFC1992,vidana16}. 
Our measurements confirm the results obtained from other experiments performed at SATURNE~\cite{roy1988}. This achievement opens great opportunities to study 
the excitation of baryonic resonances in exotic projectiles that can be produced easily by fragmentation or fission reactions at the entrance of 
the FRS spectrometer \cite{Loureiro19}. This new experiment is planned for the Phase-1 of the FAIR facility and will be discussed in future works.

\section{Conclusions}

Isobaric charge-exchange reactions induced by stable tin isotopes at energies of $1A$ GeV on different targets have been investigated at GSI using the FRS spectrometer. 
The experimental setup allows us to measure the cross sections of these processes with high accuracy and to determine in coincidence 
the missing-energy spectra of the corresponding ejectiles. Thanks to the high-resolving power of the magnetic spectrometer FRS we can clearly identify
in the missing-energy spectra the quasi-elastic and inelastic components corresponding to the nuclear spin-isospin response of nucleon-hole 
and baryonic resonance excitations, respectively.

The comparison of the missing-energy spectra obtained from the measurements with different targets allows 
to investigate the competition between $\Delta$-particle-hole and nucleon-hole excitations in nuclear matter and 
also leads to a clear observation of $\Delta$-resonance excitations in target and projectile. 
Future exclusive measurements tagging the pion emitted from the decay of the $\Delta$ resonance using 
the WASA (Wide Angle Shower Apparatus) calorimeter \cite{wasa08} will help to separate 
these contributions unambiguously, providing relevant information to understand better the formation of baryonic resonances in the nuclear medium. 

In addition, this kind of analysis could also be extended to exotic projectiles in the framework of the Super-FRS experiments \cite{sfrs16}, 
providing unique opportunities to investigate the formation and decay of other baryonic resonances, 
such as the $N$(1440) and $\Delta$(1600) resonances, in very neutron-rich and exotic nuclear matter. The same experimental setup will also be used to study 
other interesting topics in nuclear physics, such as the production of light hypernuclei \cite{saito16} and $\eta '$-mesic nuclei \cite{tanaka16}.

\section{Acknowledgments}
This work has been partially supported by the Department of Education, Culture and University Organization of 
the Regional Government of Galicia under the program of postdoctoral fellowships (ED481B-2017$/$002).


\section*{References}
\begin{thebibliography}{30}
%
\bibitem{Aoust2006}
Aoust Th and Cugnon J 2006 {\it Phys. Rev. C} \textbf{74} 064607

\bibitem{JLRS2018}
Rodr\'{i}guez-S\'{a}nchez J L et al. 2018 {\it Journal of Phys.: Conf. Series} \textbf{1024} 012002

\bibitem{krebs}
Krebs H et al. 2018 {\it Phys Rev. C} \textbf{98} 014003

\bibitem{liang}
Liang H Z et al. 2018 {\it Phys Rev. C} \textbf{98} 014311

\bibitem{feng2016}
Feng Z-Q 2016 {\it Phys. Rev. C} \textbf{94} 054617

\bibitem{sen19}
Sen D 2019 {\it Int. J. Mod. Phys. D} S0218271819501220

\bibitem{drago14}
Drago A et al. 2014 {\it Phys. Rev. C} \textbf{90} 065809

\bibitem{drago16}
Drago A et al. 2016 {\it Eur. Phys. J. A} \textbf{52} 40

\bibitem{cai15}
Cai B J et al. 2015 {\it Phys. Rev. C} \textbf{92} 015802

\bibitem{Kaufman}
Kaufman S B et al. 1979 {\it Phys. Rev. C} \textbf{20} 2293

\bibitem{chiba91}
Chiba J et al. 1991 {\it Phys. Rev. Lett.} \textbf{67} 1982

\bibitem{bachelier}
Bachelier D et al. 1986 {\it Phys. Lett. B} \textbf{172} 23

\bibitem{Geissel1992}
Geissel H et al. 1992 {\it Nucl. Instr. Methods Phys. Res., Sect. B} \textbf{70} 286

\bibitem{kelic2004}
Keli\'{c} A et al. 2004 {\it Phys. Rev. C} \textbf{70} 064608

\bibitem{JL2017v0}
Rodr\'{i}guez-S\'{a}nchez J L et al. 2017 {\it Phys. Rev. C} \textbf{96} 034303

\bibitem{Tanaka18}
Tanaka Y K et al. 2018 {\it Phys. Rev. C} \textbf{97} 015202

\bibitem{roy1988}
Roy-Stephan M 1988 {\it Nucl. Phys. A} \textbf{488} 187c

\bibitem{vargas}
Vargas J et al. 2013 {\it Nucl. Instr. Methods Phys. Res., Sect. A} \textbf{707} 16

\bibitem{Horst18}
Lenske H et al. 2018 {\it Progr. Part. Nucl. Phys.} \textbf{98} 119

\bibitem{PFC1992}
Fern\'{a}ndez de C\'{o}rdoba P et al. 1992 {\it Nucl. Phys. A} \textbf{544} 793

\bibitem{vidana16}
Vida\~{n}a I et al. 2016 {\it EPJ Web of Conferences} \textbf{107} 10003

\bibitem{Loureiro19}
P\'{e}rez-Loureiro D et al. 2019 {\it Phys. Rev. C} \textbf{99} 054606

\bibitem{wasa08}
Bargholtzl Chr et al. 2008 {\it Nucl. Instr. Methods Phys. Res., Sect. A} \textbf{594} 339

\bibitem{sfrs16}
\"{A}yst\"{o} J et al. 2015 {\it JPS Conf. Proc.} \textbf{6}, 020035

\bibitem{saito16}
Saito T R et al. 2016 {\it Nucl. Phys. A} \textbf{954} 199

\bibitem{tanaka16}
Tanaka Y K et al. 2016 {\it Phys. Rev. Lett.} \textbf{117} 202501

%
\end{thebibliography}
\end{document}